\documentclass[aps,prc,twocolumn,nofootinbib]{revtex4-1}
\usepackage{amsmath,graphicx}

\usepackage[breaklinks]{hyperref}

\begin{document}
\preprint{Saclay/T-10/...}
\title{$v_4$ from ideal and viscous hydrodynamic simulations of nuclear collisions at the BNL Relativistic Heavy Ion Collider (RHIC) and the CERN Large Hadron Collider (LHC)}
\author{Matthew Luzum}
\author{Cl\'ement Gombeaud}
\author{Jean-Yves Ollitrault}
\affiliation{Institut de Physique Th\'eorique,\\
CEA, IPhT, F-91191 Gif-sur-Yvette, France\\
CNRS, URA 2306, F-91191 Gif-sur-Yvette, France} 
\date{\today}
\begin{abstract}
We compute $v_4/(v_2)^2$ in ideal and viscous hydrodynamics.  
We investigate its sensitivity to details of the hydrodynamic model and compare the results to experimental data from RHIC.  
Whereas $v_2$ has a significant sensitivity only to initial eccentricity and viscosity while being insensitive to freeze out temperature, we find that $v_4/(v_2)^2$ is quite insensitive to initial eccentricity. On the other hand, it can still be sensitive to shear viscosity in addition to freeze out temperature, although viscous effects do not universally increase $v_4/(v_2)^2$ as originally predicted.   Consistent with data, we find no dependence on particle species.
%
We also make a prediction for $v_4/(v_2)^2$ in heavy ion collisions at the LHC.
\end{abstract}
\maketitle
\section{Introduction}
Much can be learned about the bulk properties of the medium created in relativistic heavy ion collisions by studying the azimuthal distribution of emitted particles.   Typically, the dependence on the azimuthal angle $\phi$ with respect to the collision plane is written as a Fourier series
\begin{equation}
E\frac{d^3 N}{d^3 {\bf p}}
= v_0 \left[1 + \sum_{n = 1}^\infty 2 v_n\, \cos(n\, \phi) \right].
\end{equation}
In a collision between identical nuclei at midrapidity, the sine
terms and all odd terms are negligible when averaged over many events.
There has been much study of the elliptic flow coefficient, $v_2$~\cite{Voloshin:2008dg}, which has been an important observable for determining, e.g.,  the viscosity of the medium~\cite{Luzum:2008cw,Masui:2009pw,Drescher:2007cd,Heinz:2009cv,Bozek:2009dw}.  Less studied, however is the next harmonic $v_4$~\cite{Kolb:2003zi,Kolb:2004gi}.  In this article we use viscous and ideal hydrodynamic simulations to determine what $v_4$ can tell us about the heavy ion collision fireball created at the Relativistic Heavy Ion Collider (RHIC), and to make predictions for planned heavy ion collisions at the Large Hadron Collider (LHC).

For details of the hydrodynamic model used, the reader is directed to Ref.~\cite{Luzum:2008cw}.  As a brief summary, we solve 2+1 dimensional conformal viscous hydrodynamic equations (with zero bulk viscosity) using both Glauber- and CGC-type initial conditions and a Cooper-Frye freeze out routine followed by resonance feed down.  All calculations use the same parameters as those that gave the best fit to RHIC data~\cite{Luzum:2008cw} (see Table \ref{tab:par}), except where specified.
\begin{table}
%
%
%
%
\begin{center}
\begin{tabular}{|c|c|c|c|c|}
\hline
Init. cond. & 
$\eta/s$ & 
$T_i$ [GeV] & 
$T_f$ [GeV] &
$\tau_0$ [fm/c] \\
\hline
Glauber& $10^{-4}$ & 0.340 & 0.14&1\\
{\bf Glauber}& {\bf 0.08} & {\bf 0.333} & {\bf 0.14} & {\bf 1}\\
Glauber& 0.16 & 0.327 & 0.14&1\\
CGC& $10^{-4}$ & 0.310 & 0.14&1\\
{\bf CGC}& {\bf 0.16} & {\bf 0.299} & {\bf 0.14} & {\bf 1}\\
%
%
%
\hline
\end{tabular}
\end{center}
\caption{\label{tab}Parameters used for the viscous hydrodynamics
simulations except as indicated in each relevant section. The bold faced lines give the best fit to $v_2$, $\langle p_t\rangle$, and multiplicity~\cite{Luzum:2008cw}.}
\label{tab:par}
\end{table}
\section{Review}
First recall the simplest prediction for $v_4$~\cite{Borghini:2005kd} that predicts $v_4 = 0.5 (v_2)^2$ for a boosted thermal freeze out distribution.
To summarize, in the low temperature limit, one can solve the zero-viscosity Cooper-Frye freeze out formula with Boltzmann statistics
\begin{equation}
\label{CF}
E\frac{d^3 N}{d^3 {\bf p}}  
\propto \int p_{\mu} d\Sigma^\mu \exp \left(- \frac {p_\mu u^\mu} {T} \right)
\end{equation}
by performing a saddle point approximation.   At large transverse momentum (i.e., where the minimum of $p\cdot u >  m$), the dominant part of the integral comes from the part of the freeze out surface where the fluid velocity is parallel to the momentum of the emitted particle, and is at a maximum.  We can decompose this maximum fluid 4-velocity magnitude $u_{\rm max}$ at each angle $\phi$ in a Fourier series
\begin{equation}
u_{\rm max}(\phi) = U(1 + 2V_2 \cos(2\phi) + 2V_4 \cos(4\phi) + \ldots )\, .
\end{equation}
Plugging this into Eq.~\eqref{CF} in a saddle point approximation and expanding to leading order in $V_2$ and $V_4$, one obtains
\begin{align}
v_2(p_t)=&\frac{V_2 U}{T}\left(p_t-m_t v\right)\\
v_4(p_t)=&\frac{1}{2}\frac{(V_2 U)^2}{T^2}\left(p_t-m_t v\right)^2+\frac{V_4 U}{T}\left(p_t-m_t v\right) \nonumber \\
\label{v4}
=&\frac{1}{2}v_2(p_t)^2 + \frac {V_4} {V_2} v_2(p_t),
\end{align}
with $m_t=\sqrt{p_t^2+m^2}$ and  $v\equiv U/\sqrt{1+U^2}$.  At a large enough $p_t$,
the first term in Eq.~\eqref{v4} dominates, and it is in this sense that it can be said that $v_4$ is largely generated by an intrinsic elliptic flow (i.e., $V_2$) rather than a fourth order moment of the fluid flow ($V_4$).  As one relaxes the large $p_t$ limit, the ratio of $v_4$ to $(v_2)^2$ should have corrections from $1/2$ that depend on the particular collision, but behave as $1/p_t$ (or equivalently $1/v_2$).

This derivation is the motivation for studying the quantity $v_4/(v_2)^2$ rather than $v_4$ alone, and is the quantity to be studied here.

Equation \eqref{v4} is quite general and does not depend on the details of the freeze out surface or the fluid flow that produces it.  It should be noted, however, that several assumptions must be made in order to derive it:  small temperature, large transverse momentum,  small $v_2$ and $v_4$, and zero viscosity.  As these restrictions are relaxed, the behavior becomes more complicated and more sensitive to details.  Therefore, in order to fully understand the final results of the most realistic simulations as well as to make contact with previous results from ideal hydrodynamics and transport simulations, it will be useful to gain insight by starting with the simple case of vanishing viscosity and low freeze out temperature, and then adding viscosity and a realistic freeze out temperature once the generic features of the simpler case are understood.
%
%
%
%
%
\section{Review of RHIC Data}
$v_4$ has been measured at RHIC by both the STAR 
\cite{Adams:2003zg, Poskanzer:2004vd, Adams:2004bi, Abelev:2007qg, Bai:2007ky, YutingPhD} 
and PHENIX 
\cite{Masui:2005aa, Huang:2008vd, Huang:2009zzh, :2010ux} collaborations.
They both find a significant deviation of $v_4/(v_2)^2$ from a constant 1/2.  Rather, the reported value of the ratio is close to a constant value of 1 for a large range of transverse momentum ($p_t  \gtrsim 500$ MeV)~\cite{:2010ux}, centrality ($\sim$10--70\%)~\cite{YutingPhD,  :2010ux}, and for all identified particles that have been measured (pions, kaons, and protons~\cite{Huang:2008vd, Huang:2009zzh}).   It is larger for more central events, while there is no clear consensus 
for small $p_t$ and very peripheral events.  The results from PHENIX tend to be somewhat smaller than from STAR, likely indicating different sensitivities to non-flow effects, although all results are still well above 0.5.

Much of this discrepancy between prediction and experiment can be
understood by noting that both $v_4$ and $v_2$ are averaged over many
events before the ratio is taken.  Even if the ratio $v_4/(v_2)^2$
were exactly 1/2 in each event, any fluctuation event-by-event (both
at a given centrality and within a centrality bin) will tend to
increase the measured $v_4/(v_2)^2$.  In fact, using a Glauber Monte Carlo model of eccentricity fluctuations can explain most of the difference, as demonstrated in Ref.~\cite{Gombeaud:2009ye}.  With this understanding, all calculations here---since they lack fluctuations---should be understood as relating to measured data by an overall (momentum independent) multiplicative factor that depends on the centrality bin chosen.

%
%
%
%
%
%
%
\begin{figure*}
\includegraphics[width=.497\linewidth]{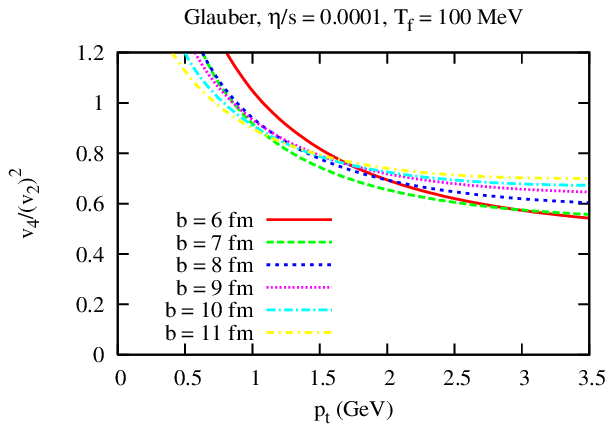}
\includegraphics[width=.497\linewidth]{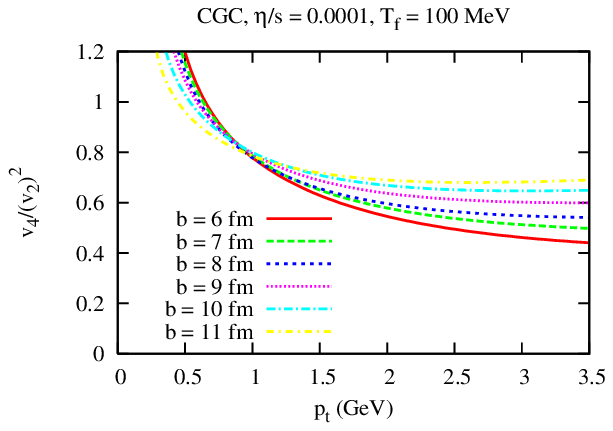}
\caption{(Color online) Pion $v_4/(v_2)^2$ from ideal hydrodynamics calculations with Glauber and CGC initial conditions and the freeze out temperature lowered to 100 MeV}
\label{fig:TF100ideal}
\end{figure*}
%
%
\section{RHIC Results}
We begin by noting that, since $v_4$ tends to be much smaller than $v_2$, numerical errors can be of greater concern.  We estimated this uncertainty in two ways.  First, we looked at the results at $b = 0$, where $v_4$ should be zero theoretically.  However, the use of a square lattice as well as the presence of numerical/rounding errors results in a calculated $v_4$ that can be as much as $\sim5$\% of the value calculated at, e.g., $b=6$ fm (or a much smaller fraction of the value for more peripheral collisions.)  Second, we reran a number of simulations with half the usual grid spacing (0.2 fm instead of 0.4 fm).  Similarly, we found that $v_4/(v_2)^2$ changed by less than 5\%--10\% as long as we only considered collisions for $b >$ 5--6 fm.  Therefore, we only report the results for $b = 6$ fm and larger.  Reliable results outside this range can be obtained with more computer time or a better hydrodynamics algorithm (as reported in Ref.~\cite{Schenke:2010nt}).

In addition to numerical uncertainties, there are several other concerns to keep in mind.  Although the results are numerically stable above $p_t \sim$0.2--0.3 GeV, they are moderately sensitive to details (e.g., small changes in parameters or the neglect of resonance feed down) up to $\sim$0.5--1 GeV, although they seem quite robust at larger $p_t$.  Of course, one must also note that in reality the hydrodynamic description should break down at some large value of $p_t$ and therefore one has less and less confidence in the results as $p_t$ increases.  However, it is not clear \textit{a priori} where exactly this break down should occur, and so it is still useful to calculate quantities even at large $p_t$.  Only comparison to data can then give information about this break down of the hydrodynamic description.

None of these concerns should change the trends and conclusions reported here.

Figure \ref{fig:TF100ideal} shows the results for pions with $\eta / s
= 0.0001$ (corresponding to ideal hydrodynamics), and all parameters as specified in Table \ref{tab:par} except a
small freeze out temperature of 100 MeV.  As found previously in ideal
hydrodynamic simulations~\cite{YutingPhD,Borghini:2005kd}, $v_4/(v_2)^2$ approaches  a constant value of roughly 1/2 at asymptotically large values of $p_t$, and increases as 1/$p_t$ for decreasing $p_t$.  In contrast to expectations and data, however, there is a definite dependence on impact parameter.  Part---but not all---of this dependence as well as the deviation from 1/2 can be understood by the fact that here $v_2$ is unphysically large, as we have turned off viscous effects.  
For example, with CGC initial conditions at $b = 11$ fm and $p_t = 3.5$ GeV, $v_2$ reaches 70\%.  $v_4/(v_2)^2$ would then be expected to reach almost 0.62 instead of 0.50 at large $p_t$ if one assumes $V_4$ of the fluid to be negligible (see Ref.~\cite{Gombeaud:2009ye}).

Figure \ref{fig:TF100ideal} also makes it apparent that, unlike $v_2$ itself, the results for $v_4/(v_2)^2$ are quite insensitive to the initial eccentricity.  Although there is a dependence on impact parameter, it is small, and there is little difference for Glauber and CGC initial conditions, despite having a very different initial eccentricity (and therefore very different $v_2$ for the same viscosity---see, e.g., Figs. 4 and 8 of Ref.~\cite{Luzum:2008cw}).  Indeed, CGC initial conditions, with a larger initial eccentricity, lead to a smaller value than Glauber while more peripheral collisions have a larger value than more central collisions, despite also having a larger eccentricity.
\begin{figure}
\includegraphics[width= \linewidth]{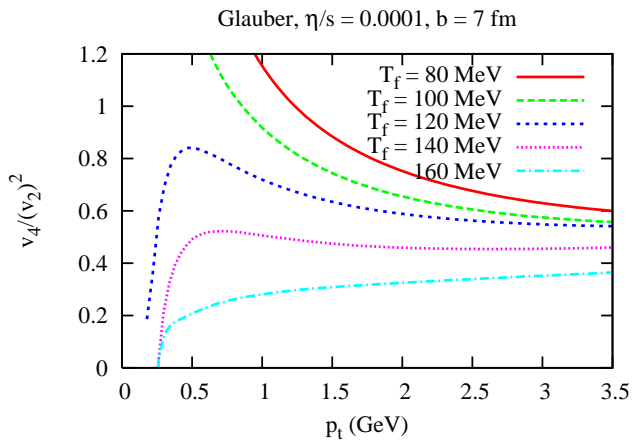}
\caption{(Color online) Pion $v_4/(v_2)^2$ for ideal hydrodynamics calculations with varying freeze out temperatures.}
\label{fig:TFscan}
\end{figure}

The effect of raising the freeze out temperature to a more realistic
value can be seen in Fig.~\ref{fig:TFscan}.
The choice of freeze out
temperature makes a significant difference to the shape of the
$v_4/(v_2)^2$ curve.  
This sensitivity to $T_f$ is perhaps not a complete surprise, since it is known from transport calculations that adjusting the treatment of freeze out does affect $v_4$~\cite{Greco:2008fs}.  
For $T_f=140$~MeV, $v_4/(v_2)^2$ is very flat as a function of $p_t$ (reminiscent of data), 
which is not the case when $T_f$ is significantly larger or smaller.  
It is interesting that this freeze out temperature is the same as gives the
best fit of charged hadron multiplicity, $\langle p_t \rangle$, and
$v_2$~\cite{Luzum:2008cw} in {\it viscous\/} hydrodynamics. 
Note that in Fig.~\ref{fig:TFscan}, viscosity has been neglected and so it corresponds to a $v_2$ that is
larger than seen experimentally.   Nevertheless, recall that the initial eccentricity---and by extension the magnitude of $v_2$---does not significantly effect  $v_4/(v_2)^2$, and so this coincidence may indeed be significant.

\begin{figure}
\includegraphics[width= \linewidth]{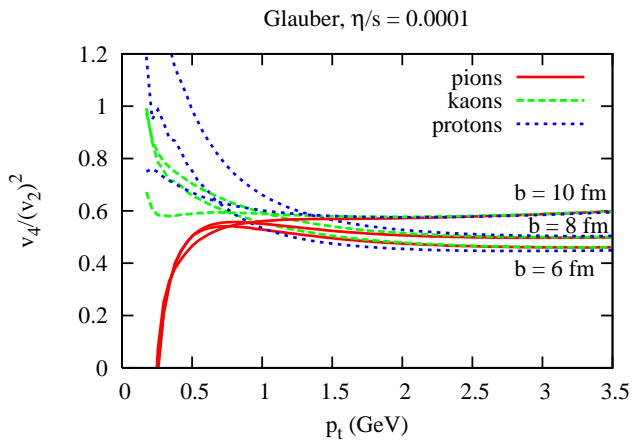}
\caption{(Color online) $v_4/(v_2)^2$ as a function of $p_t$ for identified particles from ideal hydrodynamics with Glauber initial conditions and 140 MeV freeze out temperature.}
\label{fig:140identified}
\end{figure}
\begin{figure*}
\includegraphics[width=.497\linewidth]{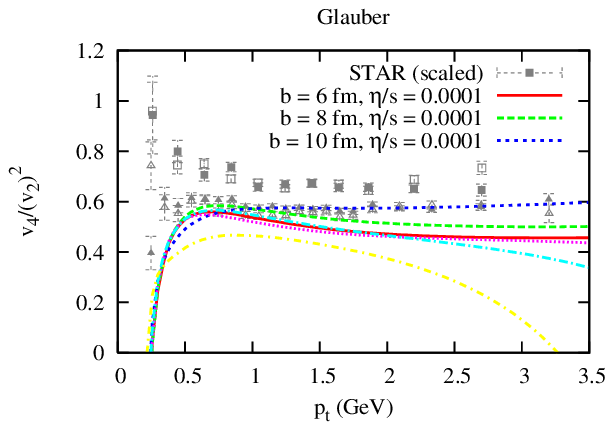}
\includegraphics[width=.497\linewidth]{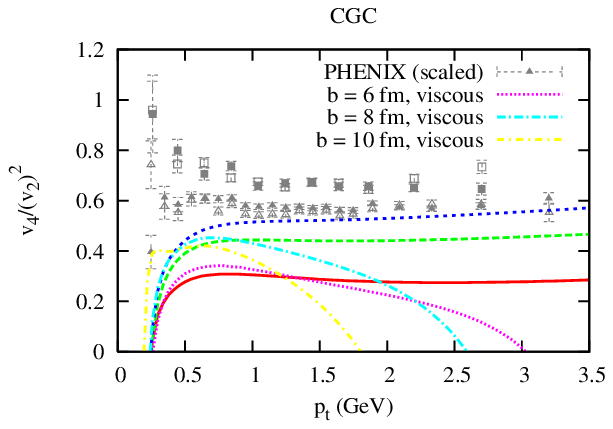}
\caption{(Color online) Charged hadron $v_4/(v_2)^2$ as a function of $p_t$  from ideal and viscous hydrodynamics with Glauber and CGC initial conditions.  Experimental results have been scaled down to account for the estimated effect of fluctuations~\cite{Gombeaud:2009ye}.  Filled and open squares are for charged hadrons at 10--20\% (with scale factor  $K=1.56$) and 40--50\% ($K=1.38$) centrality, respectively, from STAR~\cite{YutingPhD}.  Filled and open triangles are for charged hadrons at 15--20\% ($K=1.47$) and 45--50\% ($K=1.38$) centrality, respectively, from PHENIX~\cite{:2010ux}.   Viscous results correspond to the bold lines in Table \ref{tab:par}.}
\label{fig:140}
\end{figure*}

Figure \ref{fig:140identified} shows the results from ideal hydrodynamics for identified particles when the freeze out temperature is raised to 140 MeV.  Data show approximately the same value for identified particles in the measured momentum range~\cite{Huang:2008vd, Huang:2009zzh}.  This is confirmed in our calculation for $p_t\gtrsim$ 1--1.5 GeV, below which it begins to depend on the parameters of the simulation as well as increasing numerical error (especially for the heavier particles for which $v_2\sim0$).
%

Finally, Fig.~\ref{fig:140} shows the results of viscous hydrodynamic
simulations for several impact parameters, with ideal
hydrodynamics results included for comparison.  Also shown is experimental data from STAR~\cite{YutingPhD} and PHENIX~\cite{:2010ux}.
Since eccentricity fluctuations
increase $v_4/(v_2)^2$ by a factor $K\equiv\langle
\epsilon^4\rangle/\langle\epsilon^2\rangle^2$~\cite{Gombeaud:2009ye},
we have estimated $K$ in each centrality bin using the PHOBOS Glauber 
Monte Carlo~\cite{Alver:2008aq}, and scaled down the data by 
$K$. This scaling removes most of the (already small) centrality dependence for both
STAR and PHENIX in this centrality range: the residual centrality dependence is 
much smaller than in our ideal hydrodynamics results. 
The larger values of $v_4/(v_2)^2$ for
STAR as compared to PHENIX can most likely be attributed to larger nonflow
effects~\cite{Gombeaud:2009ye}. 

 It turns out that different choices for the form of the viscous
 correction $\delta f$ to the distribution function at freeze out can
 significantly change the shape of these curves (see
 Ref.~\cite{Luzum:2010ad}).  Several aspects seem to be universal for
 reasonable choices of $\delta f$, however.   In contrast to
 expectations~\cite{Bhalerao:2005mm} as well as the results at low freeze out temperature (not shown),
 viscosity does not universally increase the ration of $v_4/(v_2)^2$.
 In fact, in most cases at this realistic freeze out temperature, the
 value was found to decrease compared to ideal hydrodynamics.   Rather
 than making the agreement with data better, viscosity tends to make
 it worse.  In particular, the correction induced by viscous effects
 depends strongly on centrality: viscous effects are larger for
the smaller systems created in peripheral collisions. 
It should be noted that choosing a $\delta f$ with a weaker momentum dependence than the standard quadratic ansatz (used in Fig.~\ref{fig:140} as well as all previous viscous hydrodynamics results) can flatten these curves as well as reduce the strong impact parameter dependence (even potentially reducing it beyond that of ideal hydrodynamics), and this could in principle provide information about the dynamics of the hadron resonance gas at freeze out~\cite{Luzum:2010ad}.
\section{LHC predictions}
Using what we have learned from RHIC, it is then possible to predict what will be seen in heavy ion collisions at the LHC.  Using the same values for $\tau_0$, $T_f$, and $\eta/s$ that led to the best fit of RHIC data~\cite{Luzum:2008cw}, $T_i$ was adjusted to reproduce the expected multiplicity at top-energy Pb-Pb collisions at the LHC, as were the appropriate nuclear parameters (the same as for the LHC predictions in Ref.~\cite{Luzum:2009sb}, to which we refer the reader for details.  For reference $T_i$ is 420 MeV and 390 MeV for Glauber and CGC initial conditions, respectively.)

\begin{figure*}
\includegraphics[width=.497\linewidth]{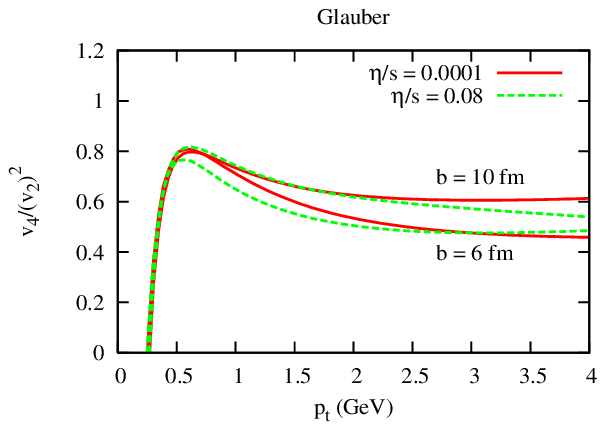}
\includegraphics[width=.497\linewidth]{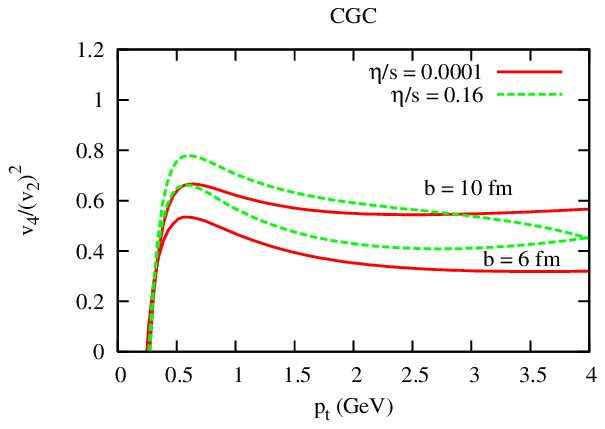}
\caption{(Color online) Charged hadron $v_4/(v_2)^2$ for 5.5 TeV Pb-Pb collisions at the LHC.  The same parameters were used as for the predictions in Ref.~\cite{Luzum:2009sb}.  The measured ratio of separately averaged quantities should be larger by a factor of $\sim$1.25--1.4, depending on the centrality bin ($K=1.34$ for a bin at 20--60\% centrality).}
\label{fig:LHC}
\end{figure*}

The results are shown in Fig.~\ref{fig:LHC}. LHC results for ideal
hydrodynamics are similar to RHIC results with a lower freeze out
temperature (see Fig.~\ref{fig:TFscan}).   Specifically, there is an increase at small $p_t$ and viscous corrections can be positive as well as negative.  However, here the viscous corrections are much smaller than for realistic RHIC simulations.  This means that---unlike for RHIC---the choice for the viscous correction to the non-equilibrium distribution function ($\delta f$) is relatively unimportant, and a comparatively robust prediction can be made, with just a correction factor expected from fluctuations put in by hand.   For example, a measurement of $v_4/(v_2)^2$ in a centrality bin of 20--60\% is expected to be larger by a factor of $\sim$1.34.  Although there is some dependence on initial conditions and impact parameter, the results are very similar to those obtained for RHIC, so it is reasonable to make the prediction that $v_4/(v_2)^2$ measured at the LHC will be similar to that measured at RHIC, except for a small increase with decreasing $p_t$.

%
%
\section{Conclusion}
%
%
%
%
In summary, we presented calculations of $v_4/(v_2)^2$ from ideal and viscous hydrodynamic simulations of heavy ion collisions at RHIC, as well as those planned at the LHC.  We confirm previous results from low $T_f$ ideal hydrodynamics and transport calculations.
This observable is found to be insensitive to initial eccentricity, but to depend strongly on the freeze-out
temperature $T_f$. In ideal hydrodynamics, the dependence on $p_t$ 
is found to be flat---as in RHIC data---only if $T_f$ is near 140~MeV, which is
precisely the value providing the best fit to $p_t$ spectra and $v_2$.    In addition, $v_4/(v_2)^2$ was found to be the same for pions, kaons, and protons---also in accord with data.  
There is an unexpected dependence on impact parameter, however, that is not seen in data.
Including viscosity using the standard freeze out scheme causes both the shape of the curve as a function of $p_t$ and the dependence on impact parameter to deviate from what is seen in data (although both aspects could potentially be fixed by using a more appropriate form for the viscous correction to the equilibrium distribution function at freeze out---see Ref.~\cite{Luzum:2010ad}). 

For the LHC, this uncertainty in the freeze out scheme is less important, and we find that $v_4/(v_2)^2$ should be similar to the value at RHIC, with a slight increase at small transverse momentum.
%
%
%
\begin{acknowledgments}
We would like to thank P.\ Bozek and W.\ Broniowski for useful discussions and P.\ Romatschke for providing the viscous hydrodynamics code.
This work was funded by ``Agence Nationale de la Recherche'' under grant
ANR-08-BLAN-0093-01.
\end{acknowledgments}


\begin{thebibliography}{99}


\bibitem{Voloshin:2008dg}
 S.~A.~Voloshin, A.~M.~Poskanzer and R.~Snellings,
 arXiv:0809.2949 [nucl-ex].

\bibitem{Luzum:2008cw}
 M.~Luzum and P.~Romatschke,
 Phys.\ Rev.\  C {\bf 78}, 034915 (2008)
 [Erratum-ibid.\  C {\bf 79}, 039903 (2009)]
 [arXiv:0804.4015 [nucl-th]].

\bibitem{Masui:2009pw}
 H.~Masui, J.~Y.~Ollitrault, R.~Snellings and A.~Tang,
 Nucl.\ Phys.\  A {\bf 830}, 463C (2009)
 [arXiv:0908.0403 [nucl-ex]].

\bibitem{Drescher:2007cd}
 H.~J.~Drescher, A.~Dumitru, C.~Gombeaud and J.~Y.~Ollitrault,
 Phys.\ Rev.\  C {\bf 76}, 024905 (2007)
 [arXiv:0704.3553 [nucl-th]].

\bibitem{Heinz:2009cv}
 U.~W.~Heinz, J.~S.~Moreland and H.~Song,
 Phys.\ Rev.\  C {\bf 80}, 061901(R) (2009)
 [arXiv:0908.2617 [nucl-th]].

\bibitem{Bozek:2009dw}
  P.~Bozek,
  Phys.\ Rev.\  C {\bf 81}, 034909 (2010)
  [arXiv:0911.2397 [nucl-th]].


\bibitem{Kolb:2003zi}
 P.~F.~Kolb,
 Phys.\ Rev.\  C {\bf 68}, 031902(R) (2003)
 [arXiv:nucl-th/0306081].

\bibitem{Kolb:2004gi}
 P.~F.~Kolb, L.~W.~Chen, V.~Greco and C.~M.~Ko,
 Phys.\ Rev.\  C {\bf 69}, 051901(R) (2004)
 [arXiv:nucl-th/0402049].

\bibitem{Borghini:2005kd}
 N.~Borghini and J.~Y.~Ollitrault,
 Phys.\ Lett.\  B {\bf 642}, 227 (2006)
 [arXiv:nucl-th/0506045].

\bibitem{Adams:2003zg}
 J.~Adams {\it et al.}  [STAR Collaboration],
 Phys.\ Rev.\ Lett.\  {\bf 92}, 062301 (2004)
 [arXiv:nucl-ex/0310029].

\bibitem{Poskanzer:2004vd}
 A.~M.~Poskanzer  [STAR Collaboration],
 J.\ Phys.\ G {\bf 30}, S1225 (2004)
 [arXiv:nucl-ex/0403019].

\bibitem{Adams:2004bi}
 J.~Adams {\it et al.}  [STAR Collaboration],
 Phys.\ Rev.\  C {\bf 72}, 014904 (2005)
 [arXiv:nucl-ex/0409033].

\bibitem{Abelev:2007qg}
 B.~I.~Abelev {\it et al.}  [the STAR Collaboration],
 Phys.\ Rev.\  C {\bf 75}, 054906 (2007)
 [arXiv:nucl-ex/0701010].

\bibitem{Bai:2007ky}
 Y.~Bai  [STAR Collaboration],
 J.\ Phys.\ G {\bf 34}, S903 (2007)
 [arXiv:nucl-ex/0701044].

\bibitem{YutingPhD}
Yuting Bai, PhD thesis, the University of Utrecht (2007). 

\bibitem{Masui:2005aa}
 H.~Masui  [PHENIX Collaboration],
 Nucl.\ Phys.\  A {\bf 774}, 511 (2006)
 [arXiv:nucl-ex/0510018].

\bibitem{Huang:2008vd}
 S.~Huang  [PHENIX Collaboration],
 J.\ Phys.\ G {\bf 35}, 104105 (2008)
 [arXiv:0804.4864 [nucl-ex]].

\bibitem{Huang:2009zzh}
 S.~Huang  [PHENIX Collaboration],
 J.\ Phys.\ G {\bf 36}, 064061 (2009).

\bibitem{:2010ux}
  A.~Adare {\it et al.}  [The PHENIX Collaboration],
  arXiv:1003.5586 [nucl-ex].

\bibitem{Gombeaud:2009ye}
 C.~Gombeaud and J.~Y.~Ollitrault,
 Phys.\ Rev.\  C {\bf 81}, 014901 (2010)
 [arXiv:0907.4664 [nucl-th]].


\bibitem{Schenke:2010nt}
  B.~Schenke, S.~Jeon and C.~Gale,
  arXiv:1004.1408 [hep-ph].


\bibitem{Greco:2008fs}
  V.~Greco, M.~Colonna, M.~Di Toro and G.~Ferini,
  arXiv:0811.3170 [hep-ph].

\bibitem{Alver:2008aq}
  B.~Alver, M.~Baker, C.~Loizides and P.~Steinberg,
  arXiv:0805.4411 [nucl-ex].

\bibitem{Luzum:2010ad}
  M.~Luzum and J.~Y.~Ollitrault,
  arXiv:1004.2023 [nucl-th].

\bibitem{Bhalerao:2005mm}
 R.~S.~Bhalerao, J.~P.~Blaizot, N.~Borghini and J.~Y.~Ollitrault,
 Phys.\ Lett.\  B {\bf 627}, 49 (2005)
 [arXiv:nucl-th/0508009].

\bibitem{Luzum:2009sb}
  M.~Luzum and P.~Romatschke,
  Phys.\ Rev.\ Lett.\  {\bf 103}, 262302 (2009)
  [arXiv:0901.4588 [nucl-th]].



\end{thebibliography}
\end{document}